\documentclass[useAMS,usenatbib]{mn2e}

\usepackage{epsfig, natbib, graphicx, color,amsmath, cancel, icomma, bm, amssymb}
\usepackage[switch,pagewise]{lineno}
\usepackage[hyperfootnotes=false]{hyperref}

\usepackage{eso-pic}

\AddToShipoutPictureBG*{%
  \AtPageUpperLeft{%
    \hspace{0.75\paperwidth}%
    \raisebox{-3.5\baselineskip}{%
      \makebox[0pt][l]{\textnormal{DES-2017-0280}}
}}}%

\AddToShipoutPictureBG*{%
  \AtPageUpperLeft{%
    \hspace{0.75\paperwidth}%
    \raisebox{-4.5\baselineskip}{%
      \makebox[0pt][l]{\textnormal{Fermilab-PUB-17-482-AE}}
}}}%

\input epsf 

\usepackage{color}

\newcommand{\Mpc}{\mbox{Mpc}}

\newcommand{\nn}{\nonumber}

\newcommand{\avg}[1]{\left\langle #1 \right\rangle}

\newcommand{\lkhd}{\mbox{$\cal{L}$}}

\newcommand{\be}{\begin{equation}}
\newcommand{\ee}{\end{equation}}
\newcommand{\bea}{\begin{eqnarray}}
\newcommand{\eea}{\end{eqnarray}}

\newcommand{\lcdm}{$\Lambda$CDM}

\newcommand{\hunits}{\mbox{km/s/Mpc}}
\newcommand{\obh}{\Omega_{\rm b}h^2}

\newcommand{\odmh}{\Omega_{{\rm dm}}h^2}
\newcommand{\Hinv}{H^{-1}}

\newcommand{\Omegam}{\Omega_{\rm m}}

\newcommand{\rs}{r_{\rm s}}
\newcommand{\rsfid}{r_{\rm s,fid}}
\newcommand{\lya}{Ly-$\alpha$}
\newcommand{\DM}{D_{\rm M}}
\newcommand{\redmagic}{redMaGiC}

\newcommand{\nuclear}{$d(p,\gamma)^3$He}

\newcommand{\planck}{\it Planck\rm}

\newcommand{\eV}{{\rm eV}}

\newcommand{\holicow}{H0LiCOW}
\newcommand{\shoes}{SH0ES}

\newcommand{\Ctot}{C_{{\rm tot}}}
\newcommand{\erf}{{\rm erf}}

\newcommand\blfootnote[1]{%
  \begingroup
  \renewcommand\thefootnote{}\footnote{#1}%
  \addtocounter{footnote}{-1}%
  \endgroup
}

\title[$H_0$ from DES Y1, BAO, and
$D/H$ Measurements]{Dark Energy Survey Year 1 Results: A Precise $\bm{H_0}$ Measurement from DES Y1, BAO, and
$\bm{D/H}$ Data}

\author[The Dark Energy Survey and the South Pole Telescope Collaborations]{
\parbox{\textwidth}{
\Large
T.~M.~C.~Abbott$^{1}$,
F.~B.~Abdalla$^{2,3}$,
J.~Annis$^{4}$,
K.~Bechtol$^{5}$,
B.~A.~Benson$^{4,6,7}$,
R.~A.~Bernstein$^{8}$,
G.~M.~Bernstein$^{9}$,
E.~Bertin$^{10,11}$,
D.~Brooks$^{3}$,
D.~L.~Burke$^{12,13}$,
A.~Carnero~Rosell$^{14,15}$,
M.~Carrasco~Kind$^{16,17}$,
J.~Carretero$^{18}$,
F.~J.~Castander$^{19}$,
C.~L.~Chang$^{6,7,20}$,
T.~M.~Crawford$^{6,7}$,
C.~E.~Cunha$^{12}$,
C.~B.~D'Andrea$^{9}$,
L.~N.~da Costa$^{15,14}$,
C.~Davis$^{12}$,
S.~Desai$^{21}$,
H.~T.~Diehl$^{4}$,
J.~P.~Dietrich$^{22,23}$,
P.~Doel$^{3}$,
A.~Drlica-Wagner$^{4}$,
A.~E.~Evrard$^{24,25}$,
E.~Fernandez$^{18}$,
B.~Flaugher$^{4}$,
J.~Frieman$^{4,7}$,
J.~Garc\'ia-Bellido$^{26}$,
E.~Gaztanaga$^{19}$,
D.~W.~Gerdes$^{24,25}$,
T.~Giannantonio$^{27,28,29}$,
D.~Gruen$^{12,13}$,
R.~A.~Gruendl$^{17,16}$,
J.~Gschwend$^{15,14}$,
G.~Gutierrez$^{4}$,
W.~G.~Hartley$^{3,30}$,
J.~W.~Henning$^{6,7}$,
K.~Honscheid$^{31,32}$,
B.~Hoyle$^{27,33}$,
B.~Jain$^{9}$,
D.~J.~James$^{34}$,
M.~Jarvis$^{9}$,
T.~Jeltema$^{35}$,
M.~D.~Johnson$^{17}$,
M.~W.~G.~Johnson$^{17}$,
E.~Krause$^{36}$,
K.~Kuehn$^{37}$,
S.~Kuhlmann$^{38}$,
N.~Kuropatkin$^{4}$,
O.~Lahav$^{3}$,
A.~R.~Liddle$^{39}$,
M.~Lima$^{40,15}$,
H.~Lin$^{4}$,
M.~A.~G.~Maia$^{14,15}$,
A.~Manzotti$^{41}$,
M.~March$^{9}$,
J.~L.~Marshall$^{42}$,
R.~Miquel$^{43,18}$,
J.~J.~Mohr$^{22,33,23}$,
T.~Natoli$^{44}$,
P.~Nugent$^{45}$,
R.~L.~C.~Ogando$^{15,14}$,
Y.~Park$^{46}$,
A.~A.~Plazas$^{36}$,
C.~L.~Reichardt$^{47}$,
K.~Reil$^{13}$,
A.~Roodman$^{12,13}$,
A.~J.~Ross$^{32}$,
E.~Rozo$^{46}$,
E.~S.~Rykoff$^{12,13}$,
E.~Sanchez$^{48}$,
V.~Scarpine$^{4}$,
M.~Schubnell$^{25}$,
D.~Scolnic$^{7}$,
I.~Sevilla-Noarbe$^{48}$,
M.~Smith$^{49}$,
R.~C.~Smith$^{1}$,
M.~Soares-Santos$^{4}$,
F.~Sobreira$^{50,15}$,
E.~Suchyta$^{51}$,
G.~Tarle$^{25}$,
D.~Thomas$^{52}$,
M.~A.~Troxel$^{31,32}$,
A.~R.~Walker$^{1}$,
R.~H.~Wechsler$^{13,12,53}$,
J.~Weller$^{23,33,27}$,
W.~Wester$^{4}$,
W.~L.~K.~Wu$^{7}$,
J.~Zuntz$^{39}$
\begin{center} 
(The Dark Energy Survey and the South Pole Telescope Collaborations)$^\dagger$
\end{center}
}
\vspace{0.4cm}
}

\begin{document}

\maketitle

\label{firstpage}

\begin{abstract}
We combine Dark Energy Survey Year 1 clustering and weak lensing data
with Baryon Acoustic Oscillations (BAO) and Big Bang Nucleosynthesis (BBN) 
experiments to constrain the Hubble constant. Assuming a flat
\lcdm\ model with minimal neutrino mass ($\sum m_\nu = 0.06\ \eV$)
we find $H_0=67.2^{+1.2}_{-1.0}\ \hunits$ (68\% CL).
This result is completely independent of Hubble constant measurements based on the distance
ladder, Cosmic Microwave Background (CMB) anisotropies (both temperature and polarization), 
and strong lensing constraints.  
There are now five data sets that: a) have no
shared observational systematics; and b) each constrain the Hubble constant
with a few percent level precision.  We compare these five independent measurements, 
and find that, as a set, the differences between them are significant at the $2.1\sigma$
level ($\chi^2/dof=20.1/11$, probability to exceed=4\%).  This difference is low enough
that we consider the data sets statistically consistent with each other.
The best fit Hubble constant obtained by combining all five data sets is $H_0 = 69.1^{+0.4}_{-0.6}\ \hunits$.
\end{abstract}

\section{Introduction}

The current standard model of cosmology is remarkably 
successful.\blfootnote{$^\dagger$For correspondence use des-publication-queries@fnal.gov}  
With only six free parameters, 
it can accurately describe the entire history of the Universe.  
The variety of data fit by this remarkable model includes: primoridal 
light element abundances \citep[e.g.][hereafter C16]{cookeetal16}; 
the temperature and polarization angular power spectra of the CMB 
anisotropies \citep[e.g.][]{planck15_cosmology,hennigetal17}; the distance--redshift
relation of standard candles such as Type IA supernovae (SNe) \citep[e.g.][]{betouleetal14}; 
galaxy--galaxy (gg) clustering in the late-time Universe \citep[e.g.][]{gaztanagaetal09,BAO6dF,BAOSDSSMain,alametal17}; 
the time delays of multiply imaged quasars \citep[e.g.][]{bonvinetal17}; 
and weak gravitational lensing measurements 
\citep[e.g.][]{mandelbaumetal13,alsingetal17,hildebrandtetal17,vanuitertetal17,troxeletal17,desy1kp}.

Despite its tremendous success and its remarkable simplicity, the standard model of cosmology is theoretically 
surprising. In this model, $\approx 85\%$ of the matter in the Universe 
is dark matter, detected only through its gravitational impact on observable matter.   
Additionally, the current accelerating expansion of the Universe
requires $\approx 70\%$ of the energy in the Universe
to take the form of either a cosmological constant, a dynamical field with negative pressure,
or a modification of general relativity.  While the cosmological constant is usually viewed
as the most conservative solution to this theoretical challenge,
its interpretation as a manifestation of vacuum energy leads to naive predictions that
differ from the observed value by many orders of magnitude 
\citep[][]{weinberg89}. 

In short, the standard model of cosmology has provided indirect evidence of not one but two 
distinct extensions of the standard model of particle physics.  
It is therefore reasonable to expect that any cracks in this standard cosmological
model might herald yet another surprise in our understanding of the cosmos.

One such possible crack arises from the value of the Hubble constant, i.e. the current rate of expansion of the
Universe.  The Hubble constant can be directly measured using type-IA SNe, whose luminosities are calibrated
using SNe hosted by nearby galaxies with known distances.
Alternatively, measurements of the CMB indirectly constrain the Hubble constant via its
impact on the CMB anisotropies.  Both of these measurements are remarkably precise.  Currently, 
the most precise SN measurement of the Hubble constant
is that of the \shoes\ collaboration \citep{riessetal16}, 
who report $H_0 = 73.24 \pm 1.74\ \hunits$.  This value is in excellent agreement with
that of \citet[][$H_0=74.3 \pm 1.5 \pm 2.1\ \hunits$]{freedmanetal12}, and is to be compared
to that inferred from {\it Planck} measurements assuming a flat $\Lambda$CDM
model with minimal neutrino mass, $H_0=67.3 \pm 1.0\ \hunits$ (\planck\ TT + low-$l$ only). 
These two values are discrepant at $3.0\sigma$.\footnote{Throughout this work, 
we rely exclusively on \planck\ TT + $\mbox{low-$l$}$ polarization data.  This ensures the \planck\ data 
set is independent of the SPTpol data set \citep{hennigetal17}.  
Including high-$l$ \planck\ polarization data increases 
the discrepancy between \planck\ and \shoes\ to $3.4\sigma$, as quoted in \citet{riessetal16}. However, 
\citet{planck15_cosmology} find evidence for instrumental systematics in their high-$l$ polarization spectra, 
and urge caution while interpreting features in them.}
This difference provides a strong motivation for searching for 
alternative methods of measuring the Hubble constant \citep{freedman17}.

As first highlighted by \citet{aubourgetal15}, the Baryon Acoustic Oscillation (BAO) signature
in the clustering of galaxies provides a standard ruler that enables us to determine $H_0$.
Slight density fluctuations in the early universe launched 
sound waves at the epoch of the Big Bang.  These sound waves traveled through the photon--baryon plasma
until the epoch of decoupling, at which point the waves were no longer pressure supported and stalled.  
The distance traveled by these waves before stalling --- the so-called sound horizon $r_s$ ---
can be readily computed \it a priori \rm for any set of cosmological parameters.
The overdensities due to these sound waves 
seeded galaxy formation, leading to a bump in the galaxy correlation function at distances
equal to the sound horizon $r_s$.  This bump is the so-called BAO feature.

Observationally, the BAO feature allows us to measure either the
angle spanned by the distance $\rs$ --- leading to a constraint on $\DM/\rs$ ---
or the redshift interval corresponding to two galaxies separated by a 
distance $\rs$ along the line of sight --- leading to a constraint on $c\Hinv/\rs$.
Here,  $\DM$ is the co-moving angular diameter distance
to the galaxies in question, and $H(z)$ is the Hubble expansion rate at the redshift of the observed galaxies.
In a flat \lcdm\ model, 
the Hubble rate is primarily sensitive
to the Hubble constant $H_0$ --- typically parameterized via $h$,
where $H_0 = 100 h\ \hunits$ --- and the total matter density parameter $\Omegam$.   
As an integral over the Hubble rate, these parameters also govern the behavior of
the angular diameter distance $\DM$.
Finally, the sound horizon $\rs$ depends on:  1) the mean
temperature of the CMB; 2) the 
dark matter density $\odmh$, and 3) the baryon density $\obh$.   In practice,
the precision with which the mean CMB temperature is known is already sufficiently high that we 
may ignore its observational uncertainties.

In summary, assuming the CMB temperature is known, 
the BAO observables $\DM/\rs$ and $c\Hinv/\rs$ fundamentally depend on
three key cosmological parameters only: $\Omegam$, $\obh$, and $h$.
BAO measurements at a single redshift will necessarily result in strong degeneracies between these 
parameters.  Fortunately, the sensitivity of the sound horizon $\rs$ to $\obh$ is relatively 
mild \citep[$d\ln \rs/d\ln \obh \approx 0.13$,][]{aubourgetal15}, so even
modest independent (i.e. non-BAO) constraints on $\obh$ suffice to break the $\obh$ degeneracy.  

Big Bang Nucleosynthesis (BBN) enables us to measure $\obh$ through its impact on the primordial deuterium-to-hydrogen
($D/H$) ratio.  During BBN, deuterium is burned to create $^4{\rm He}$.  The reaction rate increases with increasing
baryon density, so $D/H$ decreases monotonically with $\obh$.\footnote{Here, we follow \citet{planck15_cosmology} and
focus exclusively on $D/H$ observations because of the more difficult nature of the observations and interpretation
of other light elements, e.g. lithium \citep[for a review, see][]{Fieldsetal14}.}
The current best method for determining the primordial $D/H$ ratio relies on extremely
low-metallicity lines of sight to quasars, as determined from the quasar absorption spectrum.  Such pristine
lines of sight are unpolluted by baryonic processes in stars, so their element abundance ratios are expected
to be primordial.  Measurements of damped \lya\ systems in the quasar absorption spectra are used to infer the
$D/H$ ratio along these lines of sight, which in turn enables us to infer $\obh$.   

Even after including BBN data, a single BAO measurement will exhibit a strong $\Omegam$--$h$ degeneracy.
This degeneracy
ellipse rotates as the redshift is varied, so two BAO measurements that span a large redshift range can 
break this degeneracy.  \citet{aubourgetal15} and
\citet[][henceforth referred to as A17]{addisonetal17} combined low-redshift galaxy BAO measurements with
high-redshift \lya\ BAO data to arrive at a measurement of $h$.
A17 found $H_0=67.4 \pm 1.3\ \hunits$, though the authors also note
that there is an $\approx 2\sigma$ difference between the galaxy and \lya\ BAO 
 measurements.\footnote{We quote the $H_0$ value obtained
by the mean of the two values reported in A17, the more recent of the two analyses.  The two values in A17
differ on the adopted value for the $d(p,\gamma)^3\mbox{He}$ reaction rate in the BBN calculation.  
We also adopted the larger of the two error bars quoted in A17.}

In this work, we break the $\Omegam$--$h$ degeneracy of the galaxy BAO+BBN measurement
with clustering and weak lensing data from the Dark Energy Survey (DES) Year 1 data set.   
In \citet{desy1kp}, we have shown
that our analysis of the DES Y1 data results in the  most accurate and precise constraints on
the total matter density $\Omegam$ from any lensing analysis to date.  
In combination with galaxy BAO measurements and
BBN constraints derived from $D/H$ observations, we derive remarkably tight constraints on 
the Hubble rate that are independent of both CMB anisotropies and local supernova measurements.
Throughout this work we adopt $3\sigma$ (0.27\%) as the threshold
for ``evidence of tension'', and the usual $5\sigma$ ($5.7\times 10^{-7}$) 
threshold for ``definitive evidence of tension'',
though we recognize these thresholds are necessarily subjective.


\section{Analysis}

Our analysis relies on four sets of data:
\begin{enumerate}
\item The COBE/FIRAS measurements of the temperature of the CMB \citep{fixsen09}
\item Galaxy BAO measurements from a variety of spectroscopic surveys.  
\item Observational estimates of the primordial $D/H$ ratio.
\item Tomographic shear, galaxy-galaxy lensing (gg-lensing), and galaxy-galaxy clustering (gg-clustering)
data on linear scales measured in the DES Y1 data set.
\end{enumerate}

Our BAO constraints are taken directly from the constraints derived from the 6dF galaxy survey \citep{BAO6dF}, 
the SDSS Data Release 7 Main Galaxy sample \citep{BAOSDSSMain}, and the BOSS Data Release 12 \citep{BAOBoss}.
The 6dF and SDSS Main analyses were based on the monopole of the anisotropic galaxy correlation function,
and therefore do not constrain $\DM/\rs$ and $c\Hinv/\rs$ individually; rather, they constrain
the combination $D_{\rm V} = [\DM^2cz\Hinv]^{1/3}$.  Our BAO priors are listed in Table~\ref{tab:priors}.  

Our BBN priors are taken from the recent analysis by C16. Adopting
the CMB temperature of \citet{fixsen09}, C16 reports two separate constraints on $\obh$:
one obtained using a theoretical calculation for the \nuclear\ reaction rate, and one obtained using
experimental constraints for the same rate.  The two results are discrepant at $3.5\sigma$.
We adopt a conservative prior that places the central value of
$\obh$ halfway between the two values reported in C16.  The corresponding uncertainty
is set to half the difference between the two results.  Our BBN prior is reported in Table~\ref{tab:priors}.
We note that because of the mild sensitivity of the sound horizon $r_s$ to the baryon density $\Omega_b h^2$,
even a perfect measurement of $\Omega_b h^2$ would not improve the posterior of our Hubble constant
measurement in any appreciable way. 

\begin{table*}
\centering
\begin{tabular}{| l | l |}
  \hline
Prior or Data Set & Citation  \\
\hline
$D_{\rm V}(z=0.106)/\rs = 3.047 \pm 0.137$ & \citet{BAO6dF} \\
$D_{\rm V}(z=0.15)/\rs = 4.480 \pm 0.168$ & \citet{BAOSDSSMain} \\
$\DM(z=0.38)\rsfid/\rs = 1512 \pm 24\ \Mpc $ & \citet{BAOBoss} \\
$\DM(z=0.51)\rsfid/\rs = 1975 \pm 30\ \Mpc$ & \citet{BAOBoss} \\
$\DM(z=0.61)\rsfid/\rs = 2307 \pm 37\ \Mpc$ & \citet{BAOBoss} \\
$H(z=0.38)\rs/\rsfid = 81.2 \pm 2.4\ \hunits$ & \citet{BAOBoss} \\
$H(z=0.51)\rs/\rsfid = 90.9 \pm 2.4\ \hunits$ & \citet{BAOBoss} \\
$H(z=0.61)\rs/\rsfid = 99.0 \pm 2.5\ \hunits$ & \citet{BAOBoss} \\
\hline
$100\obh=2.208 \pm 0.052$ & \citet{cookeetal16} \\
\hline
$T_{\rm CMB}=2.7255 \pm 0.0006\ \rm{K}$ & \citet{fixsen09} \\
\hline
\redmagic\ clustering & \citet{desy1clustering} \\
\redmagic\ shear profiles & \citet{desy1gglensing} \\
Cosmic shear & \citet{desy1cosmicshear} \\
\hline
\end{tabular}
\caption{
BAO and BBN priors, and DES data sets used in this analysis.  The BOSS BAO priors report the
comoving angular distance and Hubble expansion relative to a fiducial sound horizon 
$\rsfid=147.78\ \Mpc$.  In practice, our analysis uses the full covariance matrix
for the BAO measurements quoted above as reported in \citet{alametal17} Table 8.
The parameter $D_{\rm V}(z)$ is defined via $ \equiv [\DM^2c\Hinv]^{1/3}$.
}
\label{tab:priors}
\end{table*}

Finally, we use the likelihood framework described in \citet{krauseetal17} to analyze the 
clustering of \redmagic\ galaxies \citep[][]{rozoetal16,desy1clustering}, the
shear profile around \redmagic\ galaxies \citep{desy1gglensing}, and the tomographic cosmic
shear signal in the DES Y1 data \citep{desy1cosmicshear}.  The shear profile and cosmic shear analyses
rely on the shape catalogs described in \citet{desy1shearcatalogs}, and the photometric redshift
analyses in \citet{desy1photozs}.  The latter include extensive validation of photometric
redshift uncertainties via cross-correlation 
methods \citep[][Cawthon et al. in prep]{gattietal17,davisetal17}. 
We refer the reader to these papers for a detailed
description of the likelihood, data vectors, and robustness and systematics checks of the
DES data. The entire framework was tested in simulations as described in MacCrann et al. (in preparation).
The DES priors employed and the
corresponding DES posteriors are presented in \citet{desy1kp}. 
Both the BBN and DES analyses were performed blind, with all analyses choices fixed
prior to revealing cosmological constraints \citep{desy1kp,cookeetal16}.  
There are also no parameter or configuration choices
made by us when performing this analysis: we are simply combining BBN, BAO, and DES data
as published.


\section{Results and Consistency with External Data Sets}

Unless otherwise noted, we adopt a flat \lcdm\ model with neutrino masses fixed at
their minimal value of $\sum m_\nu = 0.06\ \eV$, as determined from neutrino
oscillation experiments \citep[see][for reviews]{lesgourguespastor06,rpp14}. 
$N_{\mathrm{eff}}$ is also held at its expected
value $N_{{\rm eff}} = 3.046$.  This is contrary to what was done in
\cite{desy1kp}, where the neutrino mass was allowed to float by default.
Our goal here is to measure the Hubble rate with a combined DES+BAO+BBN analysis, 
and explore consistency in measurements of the Hubble constant within the context
of this maximally restrictive cosmological model.  We will, however,
demonstrate that letting the neutrino mass float has a minimal impact on our measurement
of the Hubble constant.

Unless otherwise noted, consistency between two data sets is evaluated as follows.  Let $p$ be the vector of
model parameters shared between two experiments $A$ and $B$.  We take $A$ and $B$ 
to be consistent with one another if the hypothesis $p_A-p_B=0$ is acceptable. Specifically,
for mutually independent experiments we calculate
\be 
\chi^2 = (p_A-p_B)^{{\rm T}} \Ctot^{-1} (p_A-p_B)
\ee 
and compute the probability to exceed the observed value assuming the number of degrees of freedom
is equal to the number of shared parameters. 
In the above expression, $\Ctot = C_A + C_B$ is the expected
variance of the random variable $p_A-p_B$, with $C_A$ and $C_B$ being the covariance matrix of
the shared cosmological parameters.  Both matrices are marginalized over any additional parameters
exclusive to each data set.  We evaluate the Probability-To-Exceed (PTE) $P_{\chi^2}$ of the recovered $\chi^2$ value,
and turn it into a Gaussian-$\sigma$ using the equation
\be 
P_{\chi^2}  = \erf\left( \frac{\mbox{No. of } \sigma}{\sqrt{2}} \right)
\ee 
With this definition, a probability
of $1-P_{\chi^2}=68\%$ (95\%) corresponds to $1\sigma$ ($2\sigma$) difference.
As a reminder, we have adopted $3\sigma$ difference (PTE=0.27\%) 
as our threshold for ``evidence of tension,'' and
$5\sigma$ ($\mbox{PTE}=5.96\times 10^{-7}$) 
as ``definitive evidence of tension.''

\begin{figure}
\begin{center}
\hspace*{-0.2in} \scalebox{0.6}{\includegraphics{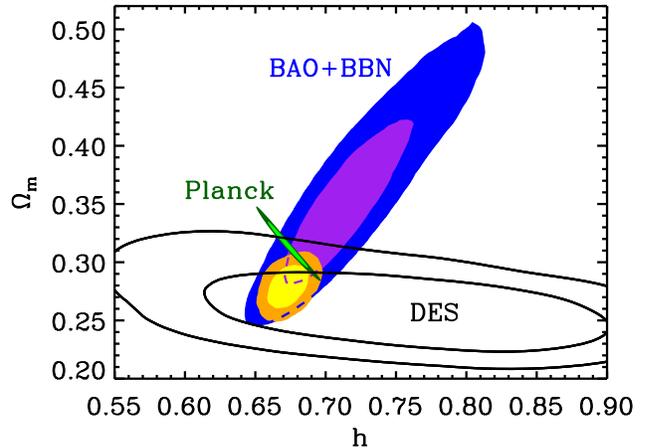}}
\caption{Constraints in the $\Omegam$--$h$ plane from the DES and BAO+BBN data as labeled.
We have adopted a definition in which $\Omegam$ includes the contribution from massive neutrinos.
All inner and outer contours enclose 68\% and 95\% of the posterior respectively.
Solid black lines show the DES $\Omega_m$--$h$ degeneracy, while the blue and purple
contours show the BAO+BBN degeneracy.  The DES+BAO+BBN contours are shown in yellow and orange. 
For reference, we have also included the corresponding contours
for the \planck\ TT+lowP data set (see text). 
}
\label{fig:Omh}
\end{center}
\end{figure}

Figure~\ref{fig:Omh} shows the $\Omegam$--$h$ degeneracy from the BAO+BBN
data (blue and purple ellipses).  
Also shown are the corresponding constraints achieved by the DES Y1 analysis (solid curves).
The two are consistent with each other at $0.6\sigma$.
A joint analysis of these data sets (yellow and orange ellipses) results in
\be
h = 0.672^{+0.012}_{-0.010}.
\ee
Throughout, we quote the most likely $h$ value, and the error bars are set by the 68\% 
contour of the posterior.
This result is in excellent agreement with and has similar precision to 
that of A17 ($h=0.674 \pm 0.013$) obtained from combining our same BAO+BBN data set with 
BAO measurements in the \lya.


We compare our posterior on $H_0$ to constraints derived from four fully independent 
datasets. These are:
\begin{itemize}
\item \planck\ measurements of CMB anisotropies as probed by the
temperature-temperature ($TT$) and low-$l$ polarization power spectra.  
The \planck\ TT+lowP data constrains $h$ when adopting a flat \lcdm\ cosmology
with minimal neutrino mass. \planck\
finds $h=0.673 \pm 0.010$ \citep{planck15_cosmology}. 
\item SPTpol has measured anisotropies in the CMB via the TE and EE angular power spectra \citep{hennigetal17}.
In our fiducial cosmological model, they find $h=0.712 \pm 0.021$.
\item The \shoes\ collaboration constrains the Hubble parameter by using type-Ia supernovae as standard candles. 
They find $h=0.732 \pm 0.017$ \citep{riessetal16}.
\item The \holicow\ collaboration constrains the Hubble parameter by measuring the time delay between
images of multiply-imaged quasars \citep{bonvinetal17}.  They find $h=0.719^{+0.024}_{-0.030}$.
\end{itemize}

\begin{figure}
\begin{center}
\hspace*{-0.3in} \scalebox{0.6}{\includegraphics{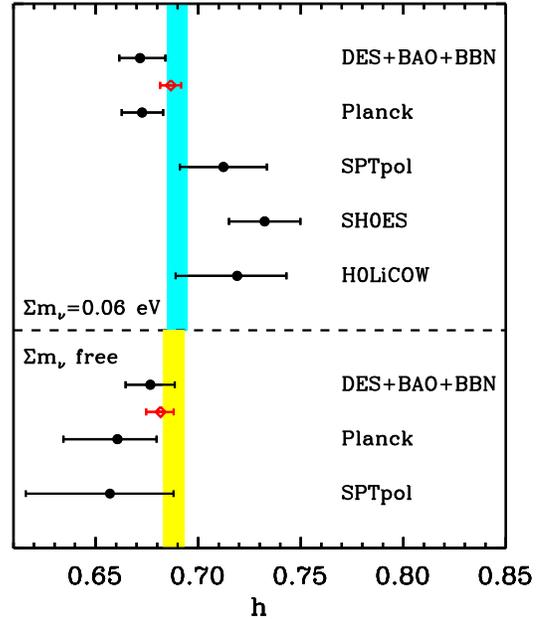}}
\caption{Posteriors on the Hubble parameter $h$ from five independent
experiments, as labeled.  
Constraints above the dashed line are obtained while
holding $\sum m_\nu$ fixed, while the constraints below the line allow the
sum of the neutrino masses to float.  In both cases, the red diamond is obtained by combining DES+BAO+BBN 
with \planck. The shift in $h$ and the greatly reduced error bars for the combined
DES+BAO+BBN and \planck\ experiments reflect the
degeneracy breaking illustrated in Figure~\ref{fig:Omh}.
The broadening and left-wards shift in the $h$ posterior from CMB experiments
reflects the degeneracy between $\sum m_\nu$ and $h$ in CMB observables (see text for
further discussion).
We emphasize that once this degeneracy is broken, all the constraints snap back into
place. The cyan and yellow bands show
the $68\%$ confidence region obtained when combining all five data sets for each of the
two analysis (fixed and free $\sum m_\nu$).  The five experiments above are 
statistically independent of each other, and share no common observational systematics.
Combining all five data sets, we arrive at $h=0.691^{+0.004}_{-0.006}$ (fixed neutrino mass)
or $h=0.689^{+0.004}_{-0.006}$ (free neutrino mass).
}
\label{fig:hcomp}
\end{center}
\end{figure}

A comparison of these various estimates of the Hubble rate and ours is shown 
in Figure~\ref{fig:hcomp}. 
All five measurements in Figure~\ref{fig:hcomp} are effectively 
statistically independent, and do not share observational systematics. 
Note in particular that the \planck\ and SPTpol data sets rely on non-overlapping
$l$-ranges in the polarization spectra with minimal sky overlap (SPTpol covers only
a small fraction of the \planck\ sky).
While the SPTpol analysis does utilize a $\tau$ prior from \planck, the posterior on
$h$ is insensitive to this prior: the constraint on $h$ is sensitive to
the relative amplitudes and positions of the acoustic peaks, not their overall amplitude.
We have explicitly verified that the SPTpol posterior on $h$ does not change
when we relax the $\tau$ prior.
Finally, while the \planck\ data does contain some information on local structures due to 
gravitational lensing, 
the volume overlap with the BAO and DES data sets is minimal, both because \planck\ is all-sky, and 
because the lensing kernel for the CMB peaks at $z\approx 2$.\footnote{In principle, we could 
remove lensing information from \planck\ by marginalizing over the so-called $A_L$
parameter.  Doing so increases the central value of the \planck\ constraint in $h$ from 0.673 to 0.689,
moving \planck\ towards the combined $h$ constraint found in this work.}

Visually, the data points in Figure~\ref{fig:hcomp} appear to be consistent with five independent
realizations of a single value.
We note that the two lowest $h$ values are the \planck\ and DES+BAO+BBN values. 
A quick look at Figure~\ref{fig:Omh} makes it obvious that when combining
these two data sets, the resulting best-fit Hubble parameter is higher that that obtained
from either data set alone, improving the agreement with the remaining data sets.
A combined DES+BAO+BBN+\planck\ analysis yields $h=0.687 \pm 0.005$, a
value higher than that of DES+BAO+BBN or \planck\ 
alone.\footnote{Combining with \planck\ improves not just the constraints on $h$, but also other cosmological
parameters, particularly $\sigma_8$ and $\Omega_m$.  Here, we focus exclusively on $h$, as this
is the key addition to the extended analysis presented in \citet{desy1kp}.}
Consistency between DES and \planck\ was established in \citet{desy1kp} using evidence
ratios.  Using the method employed in this work, we again find the two data sets
to be consistent at $1.6\sigma$.

We test for the consistency of all five data sets as follows: 
\planck\ and SPTpol provide precise measurements of $h$, $\Omega_m$, $\Omega_b$, $\sigma_8$, and $n_s$ (10 measurements).
DES+BAO+BBN measures these same parameters with the exception of $n_s$, which is not well constrained by DES.
Thus, DES+BAO+BBN adds four independent measurements.  Finally, \shoes\ and \holicow\ each measures $h$, for a total
of 16 measurements.  These are modeled using a single set of cosmological parameters (5 parameters), resulting in 11 degrees
of freedom.  We evaluate the $\chi^2$ of the best fit model to the full data vector of cosmological parameter estimates,
finding $\chi^2/dof=20.7/11$.  The probability to exceed is $4\%$, a $2.1\sigma$ difference.  We conclude that all five data sets are consistent
with each other.

We combine all five data sets to arrive at our best-fit Hubble parameter as follows.
First, we combine DES+BAO+BBN with Planck. We then 
evaluate the combined DES+BAO+BBN+\planck+SPTpol likelihood
using importance sampling (see Appendix~\ref{app:importance} for details).  
Finally, we follow a similar approach for incorporating the \shoes\ and \holicow\ 
constraints.\footnote{Since we 
do not have the \holicow\ likelihood, we have symmetrized the error bars and adopted a Gaussian likelihood.
We do not expect this approximation has a large impact on the combined posterior.}  
Combining all five data sets, we arrive at $h=0.691^{+0.004}_{-0.006}$. 
This value is consistent with earlier efforts that combined CMB, SN, and BAO oscillation
data \citep{gaztanagaetal09b}.

Of the five data sets we consider, the most discrepant $H_0$ measurement is clearly that of the
\shoes\ collaboration. As a naive estimate of the difference between \shoes\ and the remaining data 
sets, we combine all four non-\shoes\ measurements to arrive at a best estimate of the Hubble parameter
($h=0.687^{+0.005}_{-0.004}$).
The difference between this combined value and \shoes\ is $2.5\sigma$.
This value fails to satisfy our criteria for evidence of tension.
Moreover, because we have five different independent measurements,
there is an important look-elsewhere effect. Properly estimating this
effect through brute force Monte Carlo realizations of each of the five independent
data sets 
is numerically intractable.  However, we can provide a rough estimate by 
modeling the five measurements as independent Gaussian random draws of the same mean.
For each realization, we identify the random draw that is most discrepant 
relative to the remaining four values.  These four values are combined to form
a single best-estimate, and the difference between the combined result of the four
most consistent draws is compared to the remaining data point using our standard 
test for consistency.
We perform $10^5$ realizations
of this numerical experiment, and determine that the probability of finding
a difference in excess of that observed between \shoes\ and the remaining
data sets is 6\% ($1.9\sigma$).
If we instead combine the  DES+BAO+BBN with \planck\ and SPTpol, we arrive
at three independent $h$ measurements for which we can ignore the remaining cosmological
parameters.  The $\chi^2$ of these 3 independent measurements is $\chi^2/dof=7.7/2$,
corresponding to a 2.1\% probability to exceed ($2.3\sigma$).  In principle, this difference is also 
subject to a look elsewhere effect ---  we are focusing on $h$ precisely because of the \planck\ vs \shoes\ 
comparison --- 
so the significance of this difference should be slightly reduced.

We have also explored the impact of floating the sum of the neutrino masses
in our analysis.  The corresponding constraints are shown in Figure~\ref{fig:hcomp},
below the dashed line.
Opening up neutrino masses hardly impacts the recovered Hubble constant
for a DES+BAO+BBN analysis, as we would expect from the discussion in the introduction. 
Because CMB anisotropies are degenerate in $h$ and $\sum m_\nu$ --- CMB observables
are roughly constant if one increases $\sum m_\nu$ while decreasing $h$ ---
allowing $\sum m_\nu$ to float greatly increases the uncertainties in the recovered
Hubble rate from CMB experiments.  In addition, because our fiducial model corresponds
to the lower limit of $\sum m_\nu$, floating $\sum m_\nu$ necessarily shifts $h$
towards lower values, as seen in Figure~\ref{fig:hcomp}.  

The above shift is noteworthy within the broader cosmological context in that
massive neutrinos have been proposed as one way to bring the clustering amplitude
predicted from \planck\ in better agreement with low redshift measurements of $S_8=\sigma_8(\Omega_m/0.3)^{1/2}$
\citep[see e.g.][]{wymanetal14}.  The idea is simple: neutrinos don't
cluster at small scales, so increasing the fractional contribution of neutrinos to the
mass budget of the Universe decreases the predicted clustering amplitude of matter. 
However, such a shift must be accompanied by a lowering of the Hubble rate in order to
hold CMB observables fixed.  Doing so increases the difference between 
distance-ladder estimates of the Hubble constant and the DES+CMB constraints.
That is, reducing differences in $S_8$ come at the expense of increasing differences
in $H_0$. Moreover, once we combine a CMB experiment with DES+BAO+BBN, the $\sum m_\nu$--$h$ degeneracy
from CMB observables is broken, and our Hubble constant constraints snap back into place.
The posterior in $h$ when combining all five data sets while letting the neutrino mass float
is $h=0.689^{+0.004}_{-0.006}$.  Neutrino masses are also forced back towards their lower
limit: our posterior on the neutrino mass is $\sum m_\nu < 0.20\ \eV$ (95\% CL).


\begin{table*}
\centering
\begin{tabular}{| l | l | l |}
  \hline
$h$ & Data Set & Citation  \\
\hline
$0.672^{+0.012}_{-0.010}$ & DES+BAO+BBN & This work \vspace{2pt} \\
$0.673\pm 0.010$	& \planck\ & \citet{planck15_cosmology} \vspace{2pt} \\
$0.712 \pm 0.021$ & SPTpol & \citet{hennigetal17} \vspace{2pt} \\
$0.732 \pm 0.017$ & \shoes & \citet{riessetal16} \vspace{2pt} \\
$0.719^{+0.024}_{-0.030}$ & \holicow\ & \citet{bonvinetal17} \vspace{2pt} \\
\hline 
$h=0.691^{+0.004}_{-0.006}$ & Combined & This work \vspace{2pt} \\
\hline
\hline
$h=0.677^{+0.012}_{-0.012}$ & DES+BAO+BBN ($\sum m_\nu$ free) & This work \vspace{2pt} \\
$h=0.689^{+0.004}_{-0.006}$ & Combined ($\sum m_\nu$ free) & This work \\
\end{tabular}
\caption{
Hubble parameter $h$ from the 5 independent data sets considered in this work,
along with the best fit estimate coming from combining all data sets.
All data sets are mutually statistically independent, and there are no shared
sources of observational systematics between them.  Our fiducial analysis
holds $\sum m_\nu = 0.06\ \eV$, but we also report results obtained by
marginalizing over $\sum m_\nu$.
}
\label{tab:hvals}
\end{table*}


\section{Discussion}

Our combined DES+BAO+BBN analysis is similar 
in spirit to that of A17.  In particular,
whereas we break the $\Omega_m$--$h$ degeneracy inherent to a BAO+BBN measurement using DES data,
they break it using \lya-BAO data to find 
$h=0.674 \pm 0.013$, in perfect agreement with the earlier result by 
\citet{aubourgetal15}.\footnote{We averaged the two reported values from Table 3 in A17,
adding in quadrature half the difference between the two central values to the statistical error bar.}
We can directly incorporate \lya-BAO in our analysis using the
\lya$\times$\lya\ measurements of \citet{bautistaetal17} and the
\lya$\times$QSO measurements of \citet{dumasetal17}.  These results are summarized
in the latter work as
\bea 
c\Hinv(z=2.40)/\rs & = & 8.94 \pm 0.22 \\
\DM(z=2.40)/\rs & = & 36.6 \pm 1.2.
\eea
The difference between these values and the galaxy BAO measurements is $2.4\sigma$,
increasing do $2.8\sigma$ when the DES data is added to the BAO.
The addition of the \lya\ data has a minimal impact on our constraints, resulting
in a posterior $h=0.674^{+0.011}_{-0.010}$.
In principle, we could also add the recent BAO result of \citet{ataetal17},
who used quasars from the eBOSS experiment to constrain the spherically
averaged distance to $z=1.52$, but the lower precision
of this early eBOSS result will have no significant impact on our results.

Our DES+BAO+BBN analysis is also qualitatively similar to the inverse distance ladder approach
presented in \citet{aubourgetal15}, though the underlying motivation for the analysis is rather
different.  In \citet{aubourgetal15}, the sound horizon scale $\rs$ was calibrated using CMB data.
With $\rs$ in hand, \citet{aubourgetal15} used BAO to measure the comoving angular diameter distance
to redshift $z=0.57$, which was in turn used to calibrate the absolute magnitude of type Ia supernova.
This, in turn, allowed \citet{aubourgetal15} to use the Joint-Lightcurve Analysis (JLA) data set 
of \cite{betouleetal14} to measure the local Hubble parameter directly.

Compared to our analysis, the inverse distance ladder approach has the significant benefit of being
less model dependent: the local Hubble rate is measured directly in much
the same way as in the work from the \shoes\ collaboration, only now the
absolute magnitude calibration of the supernova
is based on BAO measurements at cosmological distances.

By contrast, while our DES+BAO+BBN analysis is clearly 
model dependent --- we have explicitly assumed
a flat \lcdm\ model with minimal neutrino mass --- the resulting constraint on $h$ is completely
independent of both CMB anisotropies and supernova data.  
Consequently, relative to the inverse distance-ladder, we view our analysis as a cleaner test of 
observational systematics within the specific context of a flat $\Lambda$CDM model.

Broadly speaking, our results and conclusions mirror and update those of \citet{bennettetal14}, who
pursued a similar examination to that of this work.  Like us, they find no significant evidence
of tension in Hubble constant measurements, reaching a consensus value from WMAP, BAO, and SN
data of $H_0 = 69.6 \pm 0.7\ \hunits$.  This is to be compared to our own result
of $H_0 = 69.1^{+0.4}_{-0.6}\ \hunits$.  The agreement between the two values is remarkable,
particularly given the various data updates, including \planck\ 2015 results for WMAP,
the addition of SPTpol and DES data, and updated SN constraints.

As this paper was being completed, a similar paper appeared on the arXiv \citep{linishak17}.
That work compares five different estimates of $H_0$: \planck, \shoes, \holicow, and two more:
one from BAO+BBN in conjunction with supernova, and one due to a broad variety of large scale structure
measurements, including several BAO data sets, redshift space distortion analyses, cosmic shear, and
cluster abundance data.  Relative to the analysis in \citet{linishak17},  
our analysis benefits from the fact that all the 
probes we consider are clearly statistically independent and share no common systematics.  
While our conclusions
are superficially different, we agree with their basic result: the most discrepant outlier
in our collection of $H_0$ measurements is the local $H_0$ measurement from \shoes.  
Our reduced estimate of the significance of this difference incorporates the 
look-elsewhere effects present in these type of analyses.

\section{Summary}

The combination of BAO+BBN produces a tight degeneracy between $\Omegam$ and
$h$ \citep{aubourgetal15}.  
Any independent probe of $\Omegam$ can effectively break this degeneracy, enabling a direct
measurement of the Hubble parameter that is fully independent of local $H_0$ measurements and 
CMB anisotropies.  Constraints on the matter density from lensing analyses is an especially
attractive way of breaking this degeneracy: these constraints
are sensitive to dark matter via its inhomogeneities rather than through
its impact on the expansion history.  
In that sense, they enable a holistic test of the Big Bang theory
that probes not just the expanding Universe framework, but also our understanding of
density perturbations in the Universe.

We have used the recent DES Y1 data set \citep{desy1data,desy1shearcatalogs} 
to place a precise measurement
of the Hubble constant by combining it with BAO and BBN data. 
We find $H_0=67.3^{+1.1}_{-1.2}\ \hunits$.  
Our result is in $2.8\sigma$ difference with \lya\ BAO measurements, though the
combined galaxy and \lya\ BAO measurement is in good agreement with DES.  Adding \lya-BAO data to
our DES+BAO+BBN measurement has minimal impact on our results.  While our fiducial
analysis holds the sum of neutrino masses fixed, marginalizing over neutrino mass does
not significantly relax our constraint on the Hubble constant. 

We have compared our measurement of $H_0$ to four additional experimental values
of comparable precision: \planck\ TT+lowP measurements of $H_0$ assuming a flat $\Lambda$CDM
model of minimal neutrino mass; SPTpol measurements of $H_0$ in the same cosmological model;
the local supernovae-based distance ladder measurement of $H_0$ from the \shoes\ collaboration
\citep{riessetal16};
and the H0LiCOW measurement using multiply imaged quasars from \citet{bonvinetal17}.
All five measurements are mutually statistically independent of each other, 
and there are no shared observational systematics between them.
Amongst these five, the most discrepant data set is that of the \shoes\ collaboration, which
is in $2.5\sigma$ difference with the remaining four experiments.  We estimate the probability
of finding a fluctuation this large or larger in a set of five independent measurements to be 6\%, 
a $1.9\sigma$ fluctuation.  Viewed in this broader context, the $H_0$ value from the \shoes\ collaboration
does not appear to be especially problematic.

Importantly, all $H_0$ measurements used in this work are expected to improve in precision 
in the coming years.  Future CMB experiments like Advanced ACTPol \citep{debernardisetal16}, 
SPT-3G \citep{bensonetal14},
and CMB-S4 \citep{cmbs4} will survey an order of magnitude more sky area with factors of several 
lower noise than SPTpol.  By resolving the acoustic oscillations in the damping tail in the 
polarization power spectra of the CMB, these experiments will eventually surpass Planck in 
terms of their ability to constraint cosmological parameters, including $h$ \citep{gallietal14}.
Likewise, the DES survey area will more than triple while doubling the integrated
exposure per galaxy.  Future surveys like the LSST \citep{lsst09}
will further improve upon the DES five year constraints.  BAO constraints from 
eBOSS \citep{eboss16}
will increase the galaxy BAO measurements to redshifts $z\sim 1$, only to be surpassed
by DESI \citep{2016arXiv161100036D, 2016arXiv161100037D} on a few years time scale.  Local $H_0$ measurements will improve with improved distance
calibration from Gaia \citep{gaia}, and innovative techniques such as using the tip of the
red giant branch to build the distance ladder \citep{freedman17}.  
Finally, continued monitoring and improved
lens modeling techniques will further reduce the uncertainty of strong-lens estimates of $H_0$.
Together, these improvements. along with new measurements from gravitational wave events \citep{Abbottetal17}, 
will lead to ever more stringent tests of the Big Bang model and the currently standard flat \lcdm\ model
across its full 13.8 billion year history.
\\

{\it Acknowledgements:} This paper has gone through internal review by the DES collaboration.
ER is supported by DOE grant DE-SC0015975 and by the Sloan Foundation, grant FG-2016-6443. YP is
supported by DOE grant DE-SC0015975.
Funding for the DES Projects has been provided by the U.S. Department of Energy, the U.S. National Science Foundation, the Ministry of Science and Education of Spain, 
the Science and Technology Facilities Council of the United Kingdom, the Higher Education Funding Council for England, the National Center for Supercomputing 
Applications at the University of Illinois at Urbana-Champaign, the Kavli Institute of Cosmological Physics at the University of Chicago, 
the Center for Cosmology and Astro-Particle Physics at the Ohio State University,
the Mitchell Institute for Fundamental Physics and Astronomy at Texas A\&M University, Financiadora de Estudos e Projetos, 
Funda{\c c}{\~a}o Carlos Chagas Filho de Amparo {\`a} Pesquisa do Estado do Rio de Janeiro, Conselho Nacional de Desenvolvimento Cient{\'i}fico e Tecnol{\'o}gico and 
the Minist{\'e}rio da Ci{\^e}ncia, Tecnologia e Inova{\c c}{\~a}o, the Deutsche Forschungsgemeinschaft and the Collaborating Institutions in the Dark Energy Survey. 

The Collaborating Institutions are Argonne National Laboratory, the University of California at Santa Cruz, the University of Cambridge, Centro de Investigaciones Energ{\'e}ticas, 
Medioambientales y Tecnol{\'o}gicas-Madrid, the University of Chicago, University College London, the DES-Brazil Consortium, the University of Edinburgh, 
the Eidgen{\"o}ssische Technische Hochschule (ETH) Z{\"u}rich, 
Fermi National Accelerator Laboratory, the University of Illinois at Urbana-Champaign, the Institut de Ci{\`e}ncies de l'Espai (IEEC/CSIC), 
the Institut de F{\'i}sica d'Altes Energies, Lawrence Berkeley National Laboratory, the Ludwig-Maximilians Universit{\"a}t M{\"u}nchen and the associated Excellence Cluster Universe, 
the University of Michigan, the National Optical Astronomy Observatory, the University of Nottingham, The Ohio State University, the University of Pennsylvania, the University of Portsmouth, 
SLAC National Accelerator Laboratory, Stanford University, the University of Sussex, Texas A\&M University, and the OzDES Membership Consortium.

Based in part on observations at Cerro Tololo Inter-American Observatory, National Optical Astronomy Observatory, which is operated by the Association of 
Universities for Research in Astronomy (AURA) under a cooperative agreement with the National Science Foundation.

The DES data management system is supported by the National Science Foundation under Grant Numbers AST-1138766 and AST-1536171.
The DES participants from Spanish institutions are partially supported by MINECO under grants AYA2015-71825, ESP2015-66861, FPA2015-68048, SEV-2016-0588, SEV-2016-0597, and MDM-2015-0509, 
some of which include ERDF funds from the European Union. IFAE is partially funded by the CERCA program of the Generalitat de Catalunya.
Research leading to these results has received funding from the European Research
Council under the European Union's Seventh Framework Program (FP7/2007-2013) including ERC grant agreements 240672, 291329, and 306478.
We  acknowledge support from the Australian Research Council Centre of Excellence for All-sky Astrophysics (CAASTRO), through project number CE110001020.

The South Pole Telescope program is supported by the National Science Foundation through grant PLR-1248097. Partial support is also provided by the NSF Physics Frontier Center grant PHY-0114422 to the Kavli Institute of Cosmological Physics at the University of Chicago, the Kavli Foundation, and the Gordon and Betty Moore Foundation through Grant GBMF\#947 to the University of Chicago.

This manuscript has been authored by Fermi Research Alliance, LLC under Contract No. DE-AC02-07CH11359 with the U.S. Department of Energy, Office of Science, Office of High Energy Physics. The United States Government retains and the publisher, by accepting the article for publication, acknowledges that the United States Government retains a non-exclusive, paid-up, irrevocable, world-wide license to publish or reproduce the published form of this manuscript, or allow others to do so, for United States Government purposes.

\newcommand\AAA{{A\& A}}
\newcommand\PhysRep{{Physics Reports}}
\newcommand\apj{{ApJ}}
\newcommand\PhysRevD[3]{ {Phys. Rev. D}} 
\newcommand\prd[3]{ {Phys. Rev. D}} 
\newcommand\jcap[3]{{JCAP}} 
\newcommand\PhysRevLet[3]{ {Phys. Rev. Letters} }
\newcommand\mnras{{MNRAS}}
\newcommand\PhysLet{{Physics Letters}}
\newcommand\AJ{{AJ}}
\newcommand\aap{ {A \& A}}
\newcommand\apjl{{ApJ Letters}}
\newcommand\aph{astro-ph/}
\newcommand\AREVAA{{Ann. Rev. A.\& A.}}

\bibliographystyle{mn2e}
\bibliography{mybib}

\appendix

\section{Importance Sampling with Nuisance Parameters}
\label{app:importance}

The SPTpol likelihood was written as a \texttt{CosmoMC} \citep{Lewis:2002ah} module, 
whereas the DES likelihood was written as a \texttt{CosmoSIS} \citep{Zuntz:2014csq} module.
This difference makes it difficult to run a combined chain.  Consequently, we rely on importance sampling, 
evaluating the SPTpol likelihood at each of the links of the DES+BAO+BBN+\planck\ chains.  
However, the SPTpol likelihood includes
several nuisance parameters, including two which are not prior dominated: 
$A_{80}^{\mathrm{EE}}$, the EE dust amplitude; and $D_{3000}^\mathrm{PS_{EE}}$, the EE Poisson foreground amplitude.  
One must correctly account for these nuisance
parameters in the calculation.  We describe how we do so here.

Consider two experiments $A$ and $B$.  The two experiments share a set of parameters $p$, 
but each experiment additionally
contains a set of nuisance parameters exclusive to itself, namely $q_A$ and $q_B$.  
Given an arbitrary function $f(p, q_A, q_B)$, we wish to be
able to evaluate
\bea
\avg{f} & = & \int dp dq_A dq_B\   \lkhd_A(p,q_A)\lkhd_B(p,q_B) \nonumber \\
	    &   & \quad\quad \times P_0(p)P_0(q_A)P_0(q_B) f(p,q_A,q_B),
\eea
Where $\lkhd_X$ is the likelihood for experiment $X$ and $P_0$ represents the priors for different parameter sets. We assume here that the experiments are independent of each other, and that the priors on $p$, $q_A$, and $q_B$ are separable.

We wish to importance sample MCMC results from experiment A using the likelihood from experiment B. 
In order to efficiently sample the parameter space spanned by $q_B$, we multiply and divide the integrand by 
$G(q_B)$ where $G$ is a probability distribution chosen to be
wider than the posterior of $q_B$ (as estimated from the chains of experiment $B$ alone).  We can rewrite
the above expression as
\bea
\avg{f} & = & \int dpdq_Adq_B\ \left[ \lkhd_A(p,q_A)P_0(p)P_0(q_A)P_0(q_B) G(q_B)\right] \nn \\
		&   & \quad\quad \times \left[ \frac{\lkhd_B(p,q_B)}{G(q_B)}f(p,q_A,q_B) \right] \nn \\
		& = & \avg{\frac{\lkhd_B}{G}f}_A
\eea 
where the last expectation value refers to evaluating the expectation value of the function $f\lkhd_B/G$ over
the distribution $\lkhd_A(p,q_A)P_0(p)P_0(q_A)P_0(q_B)G(q_B)$.  Note this distribution is 
separable in $(p,q_A)$, and $q_B$.  Random draws
from $\lkhd_A(p,q_A)P_0(p)P_0(q_A)$ are given by the chain from experiment $A$, while we can readily sample from the
distribution $P_0(q_B)G(q_B)$. 
To decrease the numerical noise of the integration over the nuisance parameters,
we sample 20 different sets of $q_B$ values for each link in $p$. We found this was sufficient
to achieve good convergence, and explicitly tested using chains with both half as many points, 
and twice as many points.

In short, to importance sample the SPTpol likelihood, we first oversample the DES chain according to the weights.
For each link, we assign nuisance parameters for SPTpol by randomly drawing from the distribution
$P_0(q_B)G(q_B)$.  Each link is then assigned a weight of $\lkhd_B/G$. 

Finally, to achieve more efficient sampling of the posterior of the combined DES+BAO+BBN+\planck+SPTpol chain,
we further modified our method as follows.  First, we used the SPTpol chain to compute the parameter covariance
matrix.  We use this to define a Gaussian approximation $G_{{\rm SPT}}$ to the SPT likelihood.  This Gaussian
approximation is then included in the DES+BAO+BBN+\planck\ chain, and the assigned weight to each link
becomes $\lkhd_{{\rm SPT}}/(G\times G_{\rm SPT})$.

\section*{Affiliations}
$^{1}$ Cerro Tololo Inter-American Observatory, National Optical Astronomy Observatory, Casilla 603, La Serena, Chile\\
$^{2}$ Department of Physics and Electronics, Rhodes University, PO Box 94, Grahamstown, 6140, South Africa\\
$^{3}$ Department of Physics \& Astronomy, University College London, Gower Street, London, WC1E 6BT, UK\\
$^{4}$ Fermi National Accelerator Laboratory, P. O. Box 500, Batavia, IL 60510, USA\\
$^{5}$ LSST, 933 North Cherry Avenue, Tucson, AZ 85721, USA\\
$^{6}$ Department of Astronomy and Astrophysics, University of Chicago, Chicago, IL 60637, USA\\
$^{7}$ Kavli Institute for Cosmological Physics, University of Chicago, Chicago, IL 60637, USA\\
$^{8}$ Observatories of the Carnegie Institution of Washington, 813 Santa Barbara St., Pasadena, CA 91101, USA\\
$^{9}$ Department of Physics and Astronomy, University of Pennsylvania, Philadelphia, PA 19104, USA\\
$^{10}$ CNRS, UMR 7095, Institut d'Astrophysique de Paris, F-75014, Paris, France\\
$^{11}$ Sorbonne Universit\'es, UPMC Univ Paris 06, UMR 7095, Institut d'Astrophysique de Paris, F-75014, Paris, France\\
$^{12}$ Kavli Institute for Particle Astrophysics \& Cosmology, P. O. Box 2450, Stanford University, Stanford, CA 94305, USA\\
$^{13}$ SLAC National Accelerator Laboratory, Menlo Park, CA 94025, USA\\
$^{14}$ Observat\'orio Nacional, Rua Gal. Jos\'e Cristino 77, Rio de Janeiro, RJ - 20921-400, Brazil\\
$^{15}$ Laborat\'orio Interinstitucional de e-Astronomia - LIneA, Rua Gal. Jos\'e Cristino 77, Rio de Janeiro, RJ - 20921-400, Brazil\\
$^{16}$ Department of Astronomy, University of Illinois, 1002 W. Green Street, Urbana, IL 61801, USA\\
$^{17}$ National Center for Supercomputing Applications, 1205 West Clark St., Urbana, IL 61801, USA\\
$^{18}$ Institut de F\'{\i}sica d'Altes Energies (IFAE), The Barcelona Institute of Science and Technology, Campus UAB, 08193 Bellaterra (Barcelona) Spain\\
$^{19}$ Institute of Space Sciences, IEEC-CSIC, Campus UAB, Carrer de Can Magrans, s/n,  08193 Barcelona, Spain\\
$^{20}$ High Energy Physics Division, Argonne National Laboratory, 9700 S. Cass Avenue, Argonne, IL 60439, USA\\
$^{21}$ Department of Physics, IIT Hyderabad, Kandi, Telangana 502285, India\\
$^{22}$ Faculty of Physics, Ludwig-Maximilians-Universit\"at, Scheinerstr. 1, 81679 Munich, Germany\\
$^{23}$ Excellence Cluster Universe, Boltzmannstr.\ 2, 85748 Garching, Germany\\
$^{24}$ Department of Astronomy, University of Michigan, Ann Arbor, MI 48109, USA\\
$^{25}$ Department of Physics, University of Michigan, Ann Arbor, MI 48109, USA\\
$^{26}$ Instituto de Fisica Teorica UAM/CSIC, Universidad Autonoma de Madrid, 28049 Madrid, Spain\\
$^{27}$ Universit\"ats-Sternwarte, Fakult\"at f\"ur Physik, Ludwig-Maximilians Universit\"at M\"unchen, Scheinerstr. 1, 81679 M\"unchen, Germany\\
$^{28}$ Institute of Astronomy, University of Cambridge, Madingley Road, Cambridge CB3 0HA, UK\\
$^{29}$ Kavli Institute for Cosmology, University of Cambridge, Madingley Road, Cambridge CB3 0HA, UK\\
$^{30}$ Department of Physics, ETH Zurich, Wolfgang-Pauli-Strasse 16, CH-8093 Zurich, Switzerland\\
$^{31}$ Department of Physics, The Ohio State University, Columbus, OH 43210, USA\\
$^{32}$ Center for Cosmology and Astro-Particle Physics, The Ohio State University, Columbus, OH 43210, USA\\
$^{33}$ Max Planck Institute for Extraterrestrial Physics, Giessenbachstrasse, 85748 Garching, Germany\\
$^{34}$ Astronomy Department, University of Washington, Box 351580, Seattle, WA 98195, USA\\
$^{35}$ Santa Cruz Institute for Particle Physics, Santa Cruz, CA 95064, USA\\
$^{36}$ Jet Propulsion Laboratory, California Institute of Technology, 4800 Oak Grove Dr., Pasadena, CA 91109, USA\\
$^{37}$ Australian Astronomical Observatory, North Ryde, NSW 2113, Australia\\
$^{38}$ Argonne National Laboratory, 9700 South Cass Avenue, Lemont, IL 60439, USA\\
$^{39}$ Institute for Astronomy, University of Edinburgh, Edinburgh EH9 3HJ, UK\\
$^{40}$ Departamento de F\'isica Matem\'atica, Instituto de F\'isica, Universidade de S\~ao Paulo, CP 66318, S\~ao Paulo, SP, 05314-970, Brazil\\
$^{41}$ Institut d'Astrophysique de Paris, F-75014, Paris, France\\
$^{42}$ George P. and Cynthia Woods Mitchell Institute for Fundamental Physics and Astronomy, and Department of Physics and Astronomy, Texas A\&M University, College Station, TX 77843,  USA\\
$^{43}$ Instituci\'o Catalana de Recerca i Estudis Avan\c{c}ats, E-08010 Barcelona, Spain\\
$^{44}$ Dunlap Institute for Astronomy and Astrophysics, University of Toronto, 50 St George St, Toronto, ON, M5S 3H4, Canada\\
$^{45}$ Lawrence Berkeley National Laboratory, 1 Cyclotron Road, Berkeley, CA 94720, USA\\
$^{46}$ Department of Physics, University of Arizona, Tucson, AZ 85721, USA\\
$^{47}$ School of Physics, University of Melbourne, Parkville, VIC 3010, Australia\\
$^{48}$ Centro de Investigaciones Energ\'eticas, Medioambientales y Tecnol\'ogicas (CIEMAT), Madrid, Spain\\
$^{49}$ School of Physics and Astronomy, University of Southampton,  Southampton, SO17 1BJ, UK\\
$^{50}$ Instituto de F\'isica Gleb Wataghin, Universidade Estadual de Campinas, 13083-859, Campinas, SP, Brazil\\
$^{51}$ Computer Science and Mathematics Division, Oak Ridge National Laboratory, Oak Ridge, TN 37831\\
$^{52}$ Institute of Cosmology \& Gravitation, University of Portsmouth, Portsmouth, PO1 3FX, UK\\
$^{53}$ Department of Physics, Stanford University, 382 Via Pueblo Mall, Stanford, CA 94305, USA

\label{lastpage}

\end{document}